\documentclass[twocolumn, prl]{revtex4}
\usepackage{graphicx}
\usepackage{epsfig}
\begin{document}

\title{Fermions in Optical Lattices across Feshbach Resonance}
\author{Roberto B. Diener and Tin-Lun Ho}
\affiliation{Department of Physics, The Ohio State University, Columbus, OH}
\date{\today}
\begin{abstract}
We point out that the recent experiments at ETH~\cite{Esslinger} on fermions in optical lattices, where a band insulator evolves continuously into states occupying many bands as the system is swept adiabatically across Feshbach resonance, have implications on a wide range of fundamental issues in condensed matter.  We derive the effective Hamiltonian of these systems, obtain expressions for their energies and band populations, and point out the increasing quantum entanglement of the ground state during the adiabatic sweep.  Our results also explains why only specific regions in $k$-space can be populated after the sweep as found in ref.~\cite{Esslinger}. 
\end{abstract}
\maketitle

Ever since the experimental observation of a superfluid-insulator transition of bosons in an optical lattice~\cite{Bloch}, there has been an increasing interest in quantum gases in optical lattices among different communities.  It is generally believed that all novel phases in solid state as well as many new ones specific to atomic gases can be realized in the optical lattice setting. In addition, the prospect of using these systems to process quantum information has been raised. An exciting new direction is the combination of lattice physics with Feshbach resonances, which allow one to turn particles from weakly to strongly interacting.  Not only will this increase the chance of finding new quantum phases, but also increase the speed for quantum information processing. 

The first experiment with lattice fermions across a Fesbhach resonance has recently been performed by Esslinger's group at ETH~\cite{Esslinger}.  The authors  prepare a band insulator in a deep optical lattice with two atoms per site (in different spin states) at a magnetic field where the scattering length $a_{s}$ is small and positive.  They then vary the magnetic field to drive the system adiabatically across  resonance 
and find that several bands are partially populated near and beyond the resonance. These results, simple as they may be, pose many fundamental questions.   Since the ground state near resonance cannot be a band insulator because many bands are partially populated, is it an insulator or a conductor?  If it were the former, the band insulator would then be connected $continuously$ to a new type of insulator yet to be identified, and would be only one point in a more intricate continuous family of insulators.  If it were the latter, then there must be a quantum phase transition as one approaches resonance, which remains to be discovered. It is also not clear whether this conductor is a canonical Fermi liquid,  due to pairing correlations near resonance. 

The experiments at ETH are performed in the weak tunneling limit for the lowest band with at most two fermions per well on the average. 
The very weak coupling limit is of great interest because of its relevance to solid state systems and quantum information processing. The dynamics in this limit is dominated by the energy spectrum of two fermions in a single well.  These spectra have been studied
~\cite{Wilken} for spherical and cylindrical harmonic traps by applying a pseudo-potential method 
to model the interaction, and 
for spherical traps using a two-channel model~\cite{Stoof}.  In this last reference, the parameters chosen are confined to the narrow resonance regime, while the ETH experiments\cite{Esslinger} work with wide resonances~\cite{Diener-Ho2}.  
The studies of ref.~\cite{Wilken} in effect deal with wide resonances. 
These solutions, however, are written in the relative and center of mass coordinates of the pair 
$|\Psi\rangle_{\bf R} =  |\Phi \rangle^{rel}_{\bf R} |\chi\rangle^{cm}_{\bf R}$ in a well located at ${\bf R}$, 
and are inconvenient for the introduction of tunneling between wells and for comparing with experiments. 
In recent months, a number of models have also been proposed~\cite{Carr} for fermions in optical lattices near a Feshbach resonance in the weak tunneling limit.  These models describe the resonance by a term $b^{\dagger}({\bf R}) 
a_{m \uparrow}({\bf R})a_{m\downarrow}({\bf R}) + h.c. $ which  converts a  pair of  fermions in the open channel (the $a$'s) in the same ($m$-th) band at site ${\bf R}$ into a tightly bound closed channel molecule $b({\bf R})$ (a boson).  
Any single band model, however, automatically excludes many processes allowed by symmetry which turn out to be crucial for multi-band population near resonance, and fail to describe Feshbach resonances in deep lattices at the outset. 

The goal of this paper is to point out the proper Hamiltonian for fermions in deep optical lattices across Feshbach resonances. Focusing on single well physics, we shall seek a formulation convenient for a lattice setting and derive exact results for many key properties previously~\cite{Wilken} unexplored. These results are directly related to experiments\cite{Esslinger}, force us to rethink many fundamental issues in quantum many-body physics,  bringing out many intriguing possibilities. Our formulation also applies to bosons and boson-fermion mixtures. 
 
 {\bf The Hamiltonian:}  For deep optical lattices with at most two fermions per site, the Hamiltonian is
 $H = \sum_{\bf R}^{} h_{\bf R}^{} + \hat{T}$,  where $h_{\bf R}$ is the Hamiltonian of two fermions in a deep harmonic well located at site ${\bf R}$ and $\hat{T}$ is the hopping  between wells. 
 Suppressing the site index ${\bf R}$, the single well Hamiltonian is 
$
h =  \sum^{}_{\bf m \sigma}E_{\bf m}^{} a^{\dagger}_{\bf m\sigma} a^{}_{\bf m\sigma} + \overline{\nu} b^{\dagger}b  
+  \sum_{{\bf m}, {\bf n}}^{} \alpha_{{\bf m}, {\bf n}}^{} \, \left [ a^\dagger_{{\bf m} \uparrow} 
a^\dagger_{{\bf n}\downarrow} b + {\rm h.\, c.} \right ], 
$
 where   $E_{\bf m} = \sum_{i} \hbar \omega_{i} (m_{i}+ \frac{1}{2}) $, $i=x,y,z$ are the energy levels of a harmonic oscillator with  frequencies $(\omega_{x}, \omega_{y}, \omega_{z})$, ${\bf m}=(m_{x}, m_{y}, m_{z})$;  
 $a^{\dagger}_{\bf m \sigma}$ creates a fermion in the open channel with spin $\sigma$ and  energy $E_{\bf m} $. The corresponding wavefunction is  $u_{\bf m}^{}({\bf r}) = \prod_{i=x,y,z}u^{(i)}_{m_{i}}(r_{i})$,
$u^{(i)}_{m}(s) = \left( \frac{1}{\pi d^{2}_{i}}\right)^{1/4} \frac{e^{-s^{2}/2d^{2}_{i}} }
{ \sqrt{ 2^{m} m! }} H_{m}\left( \frac{s}{d_{i}}\right)$,
where $H_{m}(x)$ is the Hermite polynomial, and $d_{i} = \sqrt{ \hbar/M\omega_{i}}$.  $b$ is the closed channel boson with its wavefunction fixed in the ground state of the harmonic well~\cite{Wilken, b-ground} . 
 $\overline{\nu}$ is the (unrenormalized) energy difference between $b$ and the open channel fermions, and  $\alpha_{{\bf m}, {\bf n}}$ is the coupling converting a pair of open channel fermions into the close channel boson $b$. 
 
Since the potential of the optical lattice is  $V({\bf r}) = V_{o} \sum_{i=x,y,z} \cos(qr_{i})$, the single particle states on a lattice are products of 1D Wannier functions in $x$, $y$, and $z$.  The tunneling term therefore takes the form $\hat{T} = - \sum_{\bf R, R' } [\sum_{\bf m, \sigma} t_{\bf m}({\bf R}, {\bf R'})a^{\dagger}_{\bf m \sigma}({\bf R})  a^{}_{\bf m \sigma}({\bf R'}) + t_b ({\bf R}, {\bf R'}) b^\dagger ({\bf R}) b({\bf R}')]$, where 
 $t_{\bf m}({\bf R}, {\bf R'})$ are the single particle hopping energies for band $\bf m$ and $t_b({\bf R}, {\bf R'})$ is the hopping energy for the closed channel molecule~\cite{T}. 
In the following, we focus on very weak tunneling ($t\rightarrow 0$), which already has very rich physics. The effect of tunneling will be discussed elsewhere. 
 
It is important to understand the functional form of $\alpha_{\bf m, n}^{}$. Recall that 
 in real space the conversion term is proportional to 
$\int D^{\dagger}({\bf R}, {\bf r}) \psi^{}_{\uparrow}({\bf R}+{\bf r}/2)
\psi^{}_{\downarrow}({\bf R}-{\bf r}/2)$, where $D$ is the tightly bound closed channel pair, 
$\psi_{\sigma}$ is the open channel fermion with spin $\sigma$,  ${\bf R}$ and ${\bf r}$ are the center of mass and relative coordinates of the fermion pair. Since $D$ is in the ground state of the harmonic well, and since 
it is of atomic size, we can write $D^{\dagger}({\bf R}, {\bf r}) =  \prod_{i} u^{(i)}_{0} (\sqrt{2} R_{i})\delta({\bf r}) b$.  Decomposing both $\psi_{\sigma}$ in the harmonic states 
$\{ u_{\bf m} \}$, we get  $\alpha_{{\bf m}, {\bf n}} =
2^{3/2}\pi^{9/4}\alpha \prod_{i} \int {\rm d} s^{}  \,u^{(i)}_{0}(\sqrt{2} s) 
u^{(i)}_{m_{i}}(s^{})u^{(i)}_{n_{i}}(s^{})$, or 
\begin{equation}
\alpha_{{\bf m}, {\bf n}}=
 \alpha \prod_{j=x,y,z}  \, {(-1)^{(m_{j}-n_{j})/2}\over \sqrt{d_j\,\,m_{j}! \,n_{j}! }}\, \Gamma({m_{j}+n_{j}+1\over 2}), 
\label{alphamn} 
\end{equation}
when $m_{j} - n_{j}$ is even, and $\alpha$ is a constant;  $\alpha_{\bf m, n}=0$ if any of the $(m_{j}-n_{j})$ is odd.
Restricting to a single band amounts to keeping only the diagonal terms in eq.(\ref{alphamn}). 
It will not describe the conversion process properly because many processes have been suppressed. 

{\bf Renormalization of short range divergences:} 
For all resonance models,  $\overline{\nu}$ contains an infinite constant since the actual closed channel pair is approximated by a point like boson $b$.  
Renormalizing $\overline{\nu}$ to a finite $\nu^{\ast}$ is a crucial step for any calculation; the finite parameters $(\nu^{\ast}, \alpha)$ then have to be converted to physical scattering quantities
such as the scattering length $a_{s}$ and the effective range $r_{o}$ in free space in order to make contact with experiments.  To illustrate the renormalization procedure, we consider cylindrical traps with frequency
$\omega_{x}=\omega_{y}=\omega_{\perp}$, and  define $\gamma = \omega_{z}/\omega_{\perp}$, 
$d_{\perp}= \sqrt{\hbar/M\omega_{\perp}}$. 

To perform the renormalization, we start with the  
eigenstate $|\Psi \rangle = \left (\beta \,b^\dagger + \sum_{{\bf m}, {\bf n}} \eta^{}_{{\bf m}, {\bf n}} a^\dagger_{{\bf m} \uparrow} a^\dagger_{{\bf n} \downarrow} \right  ) |\emptyset \rangle. $
The Schrodinger equation $H|\Psi\rangle = E|\Psi\rangle$ then gives    \begin{eqnarray}
\eta_{{\bf m}, {\bf n}} &=& \beta \, {\alpha^{}_{{\bf m}, {\bf n}}\over E-E_{{\bf m}, {\bf n}}},\\
E-\overline{\nu} &=& \sum_{{\bf m}, {\bf n}} { \alpha_{{\bf m}, {\bf n}}^2 \over E-E_{{\bf m}, {\bf n}}}\\
\beta^{-2} &=&  1 + \sum_{{\bf m}, {\bf n}} { \alpha_{{\bf m}, {\bf n}}^2 \over (E-E_{{\bf m}, {\bf n}})^2 }.\label{beta-2} 
\label{energies_multi}
\end{eqnarray}
where $E_{{\bf m}, {\bf n}} = E_{{\bf m}} + E_{{\bf n}}$. 
Substituting (\ref{alphamn}) into (\ref{energies_multi}) one sees that the sum in 
eq. (\ref{energies_multi}) is divergent. The divergence can be extracted by putting a cut off $K^{\ast}$ in the sum as explained in ref.~\cite{AppendixI}, and we have 
\begin{equation}
E-\overline{\nu}  =  -C  \sum_{K=0}^{K^{\ast} } F(K),  \,\,\,\,\,\,\,  F(K) =
 \frac{  \Gamma \left( \frac{K-x}{\gamma}\right) }{ \Gamma\left(  \frac{K-x}{\gamma}+ \frac{1}{2}\right) }
\label{nu-sch} \end{equation}
where $C= \frac{ M \alpha^2 \pi^{7/2}}{2\hbar^2  d_{\perp} \sqrt{\gamma} }$,  and $x = \frac{E-E_{o}}{2\hbar\omega_{\perp}}  = \frac{  E}{2\hbar\omega_{\perp}} - 1 - \frac{\gamma}{2}$ is the excitation energy in the center of mass frame. 
For large $K^{\ast}$, the sum in (\ref{nu-sch}) is
$\sum^{K^{\ast}}_{K=0} F(K)  = W+ 2 \sqrt{\gamma K^{\ast}}$~\cite{div}, where 
$W = \sum_{K=0}^{\infty} \left[  
\frac{  \Gamma \left( \frac{K-x}{\gamma}\right) }{ \Gamma\left(  \frac{K-x}{\gamma}+ \frac{1}{2}\right) }
- \sqrt{ \frac{ \gamma}{K+1} }\right] + 
\sqrt{\gamma} \zeta\left(\frac{1}{2}\right)$, 
$\zeta(1/2)= -1.46$. We then have 
\begin{equation}
E - \nu^{\ast}= -C  W, \,\,\,\,\,\,\,\,\, \nu^{\ast} = \overline{\nu}  -  2 C \sqrt{ \gamma K^{\ast}} 
\label{nuast-sch} \end{equation}
With short range divergences removed, eq.(\ref{nuast-sch}) can be used to find the energy levels for given $\nu^{\ast}$ and $\alpha$.

{\bf Relating $(\nu^{\ast}, \alpha)$ to $(a_{s}, r_{o})$:}  For homogeneous systems, the energy of a bound pair $-|E|$ just below resonance is given by $-a^{-1}_{s} - r_{o}\kappa^2/2 + \kappa =0$, or 
$\sqrt{ - E} =   \frac{ \hbar}{a_{s} \sqrt{M}}   - \left( \frac{ r_{o} \sqrt{M}}{2\hbar} \right) E$, 
where $|E|= \hbar^2 \kappa^2/M$. This case can be recovered from eq.(\ref{nuast-sch})  by taking $\omega_{\perp} \rightarrow 0$ with  the bound state energy held fixed, which means $x\rightarrow -\infty$. Eq.(\ref{nuast-sch})  then reduces to  $\sqrt{ - E} =A (E-\nu^{\ast})$, where 
$A =C^{-1} \sqrt{\frac{\hbar^2}{2 \gamma M d_{\perp}^2} }$. 
Comparing it with the free space expression mentioned about, we have the relation 
\begin{equation} 
\frac{1}{a_{s} }=  - \frac{ \sqrt{2} \nu^{\ast} \hbar^3}{ M^{3/2} \alpha^2 \pi^{7/2} }  =  -\frac{\nu^{\ast} |r_{o}| M}{2\hbar^2}
\end{equation}
(notice that the effective range in a two-channel model is negative).  We can then rewrite eq.(\ref{nuast-sch})  in terms of physical parameters $a_s$ and $r_{o}$, 
\begin{equation}
\frac{d_{\perp}}{a_{s}} \sqrt{2\gamma}  + \frac{ \sqrt{2\gamma} |r_o| }{d_{\perp}} 
\left( x + 1 + \frac{ \gamma }{2} \right) 
= -  {W}, 
\label{final}\end{equation} 
which enables one to find the energy spectrum for any scattering length and effective range.  

{\bf Wide resonances:}  For $^{40}$K, $|r_{o}|$ can be obtained from the method in ref.~\cite{Diener-Ho2} and is found to be 2.4 nm~\cite{comment0}, while 
it is even smaller for $^{6}$Li.  For typical traps,  we then have $r_{o}/d_{\perp}\ll 1$. 
This corresponds to the so-called ``wide resonances"  where the wavelength $k^{-1}$ of the typical scattering process ($k^{-1}\sim d_{\perp}$ in our case) is much larger than the effective range $r_{o}$. The main characteristic of wide resonances is that the system will have very small closed channel component $\beta\ll 1$ 
and many properties become universal.  The former can be seen from eq. (\ref{beta-2}), which can be summed to~\cite{AppendixII}, 
\begin{equation}\label{beta-2f}
\beta^{-2}  = 1 +\sqrt{\gamma\over 2}  \frac{ d_{\perp} }{\pi |r_{o}|}  \sum_{p=0}^{\infty} 
\Phi(1, 2, \gamma p-x){\Gamma(p+1/2)\over p!}
\label{close} \end{equation}
where $\Phi(z, s,x) = \sum_{n=0}^{\infty} \frac{z^{n}}{ (n+ x)^{s}} $  is the $\Phi$-function.
The emergence of universality can be seen from (\ref{final}),  which  reduces to $\frac{d_{\perp}}{a_{s}} \sqrt{ 2\gamma} =-   {W}$ when $|r_{o}|/d_{\perp}\ll 1$.  At resonance, $a_{s}= \pm \infty$, the energy spectrum is found from $W(x)=
0$, completely independent of any microscopic details of the system.  Note also that in this limit, eq.(\ref{final}) reduces to the results of refs.~\cite{Wilken} in the corresponding regimes.

\begin{figure}
\includegraphics[width=3in]{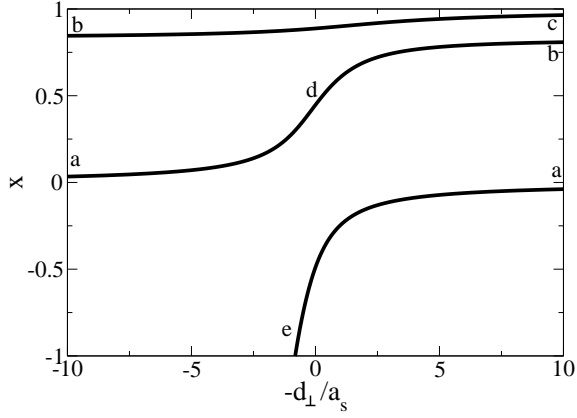}
\caption{Energy spectrum ($x = {E\over 2\hbar \omega}-1-\gamma/2$) of two fermions in a harmonic well versus inverse scattering length.  We have chosen parameters used in ref.\cite{Esslinger}, $\gamma = 5/6$, (see text).} \label{Wilkens_fig}
\end{figure}

{\bf Population of different bands:} The probability of a fermion  (with either spin) in the 
${\bf m}$-th band is 
\begin{equation}
P({\bf m}) =  \sum_{\bf n} \left| \eta_{{\bf m}, {\bf n}}\right|^2  = \sum_{\bf n} \frac{\beta^{2} \alpha^{2}_{\bf m,n}}{(E-E_{m,n})^2 },
\label{Pm1} \end{equation}
which converges for all values of the energy $E$.  In most cases, the calculation can be simplified.  For instance, for a band with indeces $(0,0,m)$, we get
$P(0,0,m)  =   {\beta^2 \gamma C\over 2\hbar \omega_\perp \pi^{3/2}} \sum_{n=0}^{\infty} 
\frac{ \Gamma \left(\frac{1}{2}(m+ n + 1)\right)^2}{n! \, m!} 
 \Phi\left(    \frac{1}{4}, 2, \gamma \left( \frac{m+n}{2}\right) -x\right)$
where the sum is restricted to values of $n$ such that $m+n$ is even ~\cite{AppendixIII}.

{\bf Our results: (I)  Energy spectrum:} In figure \ref{Wilkens_fig}, we have plotted the scaled energy $x$ as a function of  $- d_{\perp}/a_{s}$ using eq.(\ref{final})  for parameters of the ETH experiment:  $\gamma = 5/6$ and  $d_{\perp}=\sqrt{\hbar/(M\omega_{\perp})} = 63\, {\rm nm} $, $|r_{o}|/d_{\perp}= 0.04$.  Even if we set $|r_{o}|/d_{\perp}=0$ (very wide resonance), the results change only by 
a small percentage. 
The eigenstates ($|a\rangle, |b\rangle, |c\rangle$ in fig.1)  at  $- d_{\perp}/a_{s} = \pm \infty$ correspond to the lowest three eigenstates of non-interacting fermions affected by the interaction; where 
$|a\rangle=   |{\bf 0}\rangle_{rel}|{\bf 0}\rangle_{cm}$, 
$|b\rangle = |0,0,2\rangle_{rel}|{\bf 0}\rangle_{cm} $,
and $|c\rangle = 2^{-1/2}( |2,0,0\rangle_{rel} + |0,2,0\rangle_{rel}) |{\bf 0}\rangle_{cm} $. 
 States near resonance such as $|d\rangle$ in fig.1 are more complex combinations of harmonic states in the relative coordinates.  Note that $|b\rangle$ and $|c\rangle$ are stretched along $z$ and the $xy$-plane with respect to the ground state $|a\rangle$ in the relative coordinate, respectively.  It is also easy to see that in terms of the single particle basis,
both $|a\rangle$ and $|b\rangle$ can then be represented as  $\sum_{m_{z}, n_{z}}A_{m_{z}, n_{z} }| m_{z}, n_{z} \rangle $, with ${\bf m}_{\perp}={\bf n}_{\perp} = (0,0)$, where ${\bf m}_{\perp}=(m_{x}, m_{y})$.   
In particular, we have $|a\rangle = |0,0\rangle$, and 
\begin{equation}
|b\rangle = \frac{1}{2} |0,2\rangle + \sqrt{ \frac{1}{2}} |1,1\rangle + \frac{1}{2} |2,0\rangle, 
\label{entangle} \end{equation}
which is an entangled state.  Thus,  as the ground state in ref.~\cite{Esslinger} evolves adiabatically from $|a\rangle \rightarrow|d\rangle\rightarrow|b\rangle$,  it acquires {\em orbital deformation} and {\em entanglement}. 
Since the difference between $|a\rangle$ and $|b\rangle$ is their wavefunctions along $z$, there will be no change in their momentum distribution in $k_x$ and $k_y$ directions. Their differences will show up  as a spreading of the original fully filled $(0,0,0)$-band  (in $|a\rangle$) to the $(0,0,1)$ and $(0,0,2)$ bands (in state $|b\rangle$)  along $z$ as shown in figure 2\cite{Zone}.

\begin{figure}
\includegraphics[width=3in]{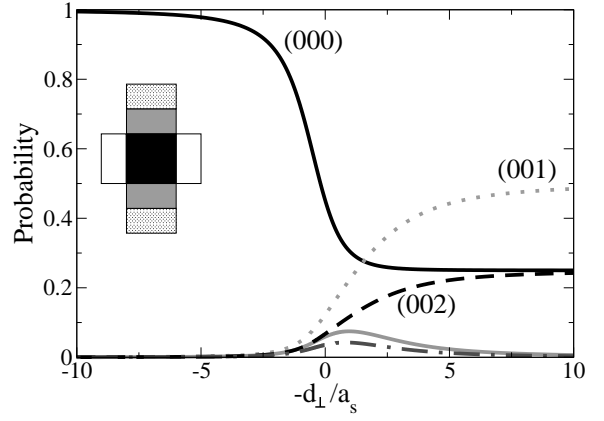}
\caption{Population of energy bands for the process $|a\rangle\rightarrow |d\rangle\rightarrow |b\rangle$ in Figure \ref{Wilkens_fig}. The black solid, dotted, and dashed curves correspond to the $(0,0,0)$, $(0,0,1)$ and $(0,0,2)$ bands, resp. The grey solid and dashed-dotted lines are for $(0,1,0)$ and $(0,2,0)$.
In $|a\rangle$, the $(0,0,0)$-band is fully filled, which occupies the entire first Brillouin Zone (black area in inset).  In state $|b\rangle$, the bands $(0,0,0)$, $(0,0,1)$, $(0,0,2)$ are $25\%, 50\%, 25\%$ occupied.  The latter two bands corresponds to the grey and dotted areas in the inset.  Higher Brillouin Zones in $xy$-directions are empty in state $|b\rangle$. 
}\label{Probabilities_fig}
\end{figure}

{\bf (II) Band populations:}  In figure 2, we have plotted the population of some of the lowest energy bands for the branch going from $a$ to $b$ at different  $- d_{\perp}/a_{s} $ using eq. (\ref{Pm1}).  All populations can be worked out analytically in the manner explained in \cite{AppendixI,AppendixII}.
As one approaches resonance starting from the band insulator $|a\rangle$, many bands are populated. The structure of the entangled state $|b\rangle$ is manifested in the 2 to 1 ratio of the population of the second and fourth bands ${\bf m} = (0,0,1)$ and $(0,0,2)$. 
Near resonance, five bands are significantly populated, which make up 85$\%$ of the probability. 
The population of the close channel molecule $\beta^2$ is less then 1$\%$ according to eq.(\ref{close}) for the process $|a\rangle \rightarrow |b\rangle$. 

{\bf Comparison with the  ETH experiments\cite{Esslinger}:}  
Our results show agreement but also discrepancies with the findings in ref.\cite{Esslinger}.
 First of all, our results predict that the momentum distribution of $|b\rangle$ will differ from that of 
 $|a\rangle$ only along $k_{z}$, with substantial population in the $(0,0,1)$ band. These are observed  in~\cite{Esslinger}.   However, the fraction of particles in the $(0,0,1)$-band in $|b\rangle$ is about 20$\%$ in ref.\cite{Esslinger}, where as the exact single well result predicts $P(0,0,1)=50\%$, (see fig. 2).   In addition, ref.\cite{Esslinger} did not observe a significant population of the $(0,0,2)$ band in $|b\rangle$, whereas our results shows $P(0,0,1)=25\%$. Recently, we learned from Michael K\"ohl that the fraction ($f$) of singly occupied sites in ref.~\cite{Esslinger} may be as large as 50$\%$.  The population of the $(0,0,1)$-band according to our result will then be $(1-f)P(0,0,1)\approx 25\%$, which agrees with \cite{Esslinger} within experimental uncertainty of number of doubly occupied sites. The absence of significant population in the $(0,0,2)$ band may be due to large tunneling at that energy, which can lead to particle loss when the band is populated. 
Another discrepancy is that the original resonance appears to be shifted in ref.~\cite{Esslinger} whereas it is essentially unshifted in the exact single well solution. 
We suspect this ``shift" may be due to the imaging process, which first lowers the barrier between wells slowly (in order to turn quasi momenta into real momenta) before a rapid turn off of the trap. In the unitarity regime, where scattering is strongest, there will certainly be a re-distribution of particles from higher bands into lower ones as the barrier is lowered, which may appear to be a shift of the resonance.   
More experiments, however, are needed to clarify the situation. We hope that our results will serve as a guide for future investigations. 

{\bf Final Remarks:}  The ETH experiment\cite{Esslinger} is yet another example of the rich and subtle physics of Feshbach resonance.  While the phenomena may appear benign at casual inspection, as explained in the introduction, they have profound implications on many-body physics.  In fact, we are in a lucky situation. Far from resonance, fermions in deep lattices can simulate nearly all important models in solid state physics by varying the barrier height, and hence the ratio $U/t$ (interaction/tunneling).  Yet, near resonance, one can have a whole host of different states whose properties are yet to be determined.  Resonance physics also makes lattice fermions a fertile ground for searching non-Fermi liquids as well as for new kind of insulators. Although we have discussed mainly about the branch labelled $a$-$b$ in fig.1, the branch $e$-$a$ is equally fascinating for one can turn a band insulator into a Bose superfluid.  The prospects are truly exciting. 

We thank Tilman Esslinger  and Michael K\"ohl for very helpful discussions.  
This work is supported by  NASA GRANT-NAG8-1765  and NSF Grant DMR-0426149 and was prepared in part at the Aspen Center for Physics.


\begin{thebibliography}{99}
\bibitem{b-ground} Since the center of mass motion of the fermion pairs is decoupled from its relative motion, it is unchanged  across resonance once it is prepared in the ground state.
\bibitem{AppendixI} 
Substituting (\ref{alphamn}) into (\ref{energies_multi}),  setting $m_{j}+n_{j}=2k_j$, and performing the sum over $m_{j}$ we arrive at $E-\overline{\nu} = \gamma C\pi^{-7/2} \, {\cal J} $, 
${\cal J} = \sum_{k_x,k_y,k_z}\, {1\over x-(k_x+k_y+\gamma k_z)}\, 
\prod_{j=x,y,z} {4^{k_j} \Gamma(k_j+{1\over 2})^2  \over (2k_j)! }$.
Using that $k!=\Gamma(k+1)$, the relation $\Gamma(2x) = 2^{2x-1}\Gamma(x)\Gamma(x+1/2)/\sqrt{\pi}$, and introducing $K=k_x+k_y$, we have 
${\cal J} = \sum_{K,k_z} {\pi^{5/2}\over x-K-\gamma k_z} {\Gamma (k_z+1/2)\over k_z!},
$
where we have used that
$\sum_{k=0}^M {\Gamma(k+1/2) \over k!}\, {\Gamma(M-k+1/2) \over (M-k)!} = \pi
$
independent of $M$.  The $K$-sum is divergent and its divergence can be extracted by imposing a cutoff $K^{\ast}$.  Finally, performing an infinite sum over $k_z$ (which is convergent) and using 
$\sum_{k_z=0}^\infty \, {\Gamma(k_z+1/2)\over k_z! (k_z-y)} = {\pi \Gamma (-y)\over \Gamma(1/2-y)}
$
we obtain eq. (\ref{nu-sch}).
\bibitem{comment0}
Given that near resonance the scattering length as a function of the magnetic field satisfies
$a_s = a_{bg} \Delta B/(B-B_0)$, the effective range satisfies
$r_0 = 2\hbar^2/(M\Delta\mu\,\Delta B\,a_{bg})$, where $\Delta \mu$ is  the difference in magnetic moment between the open and closed channels.  For $^{40}$K we have $\Delta \mu \approx 2\mu_B$, $\Delta B = 8 G$, $a_{bg} = 170 a_B$.
\bibitem{div} Writing $\sum^{K^{\ast}}_{K=0}$$(F(K)-\sqrt{\frac{\gamma}{K+1}})$$+$$\sum^{K^{\ast}}_{K=0}\sqrt{\frac{\gamma}{K+1}}$, we can set $K^{\ast}$ in the first term to infinity since the sum converges. 
 The second term can be shown easily to be $2\sqrt{\gamma K^{\ast}} + 
\sqrt{\gamma}\zeta(1/2) + 0(1/K^{\ast})$.
\bibitem{AppendixII}  Following the steps in footnote~\cite{AppendixI}, we get
$
\beta^{-2}-1 = {\gamma C \pi^{-7/2} \over 2\hbar \omega_\perp}  \sum_{K,k_z} {\pi^{5/2}\over (x-K-\gamma k_z)^2} {\Gamma (k_z+1/2)\over k_z!},\nonumber
$
which, after performing the sum in $K$, yields (\ref{beta-2f}).
\bibitem{AppendixIII} For ${\bf m}=(0,0,m)$, we set 
$n_j=2k_j$ for $j=x,y$ in eq.(\ref{Pm1}).  Performing the sum in 
eq.(\ref{Pm1}) over $k_x$, we get $
P(0,0,m)  =   {\beta^2 \gamma C\over 2\hbar \omega_\perp \pi^{3/2}} \sum_{n, K} 
\frac{ \Gamma \left(\frac{1}{2}(m+ n + 1)\right)^2}{4^K \,n! \, m!\,(x-\gamma{m+n\over 2}-K)^2} \nonumber
$
where the $n$-sum is restricted as mentioned in the text. Further summation of $K$ gives $P(0,0,m)$ in the text.
\bibitem{Zone} The $(0,0,1)$ band occupies the second and third Brillouin Zone in $z$ (but not in $xy$) direction, and $(0,0,2)$ band occupies even higher bands along $z$.
\end{thebibliography}
\end{document}